%% file: paper.tex
\documentclass[a4paper,UKenglish,cleveref,autoref,thm-restate]{lipics-v2021}

\hideLIPIcs  

\input{mymacro}

\title{On Singleton Self-Loop Removal for Termination of LCTRSs with Bit-Vector Arithmetic} 

\titlerunning{On Singleton Self-Loop Removal for Termination of BV-LCTRSs} 

\author{Ayuka Matsumi}%
{Graduate School of Informatics, Nagoya University, Japan}%
{matsumi@trs.css.i.nagoya-u.ac.jp}%
{}%
{}

\author{Naoki Nishida}%
{Graduate School of Informatics, Nagoya University, Japan}%
{nishida@i.nagoya-u.ac.jp}%
{https://orcid.org/0000-0001-8697-4970}%
{}

\author{Misaki Kojima}%
{Graduate School of Informatics, Nagoya University, Japan}%
{k-misaki@trs.css.i.nagoya-u.ac.jp}%
{https://orcid.org/0000-0001-5194-3947}%
{}

\author{Donghoon Shin}%
{Graduate School of Informatics, Nagoya University, Japan}%
{}%
{}
{}

\authorrunning{A. Matsumi, N. Nishida, M. Kojima, and D. Shin} 

\Copyright{Ayuka Matsumi, Naoki Nishida, Misaki Kojima, and Donghoon Shin} 

\ccsdesc[100]{Theory of computation~Rewrite systems} 

\keywords{%
constrained rewriting, 
dependency pair framework, 
imperative program
} 

\category{} 

\funding{This work was partially supported by JSPS KAKENHI Grant Number 18K11160.} 

\nolinenumbers 

\EventEditors{}
\EventNoEds{1}
\EventLongTitle{}
\EventShortTitle{}
\EventAcronym{}
\EventYear{}
\EventDate{}
\EventLocation{}
\EventLogo{}
\SeriesVolume{}
\ArticleNo{}

\begin{document}

\maketitle

\begin{abstract}
As for term rewrite systems, the dependency pair (DP, for short) framework with several kinds of DP processors is useful for proving termination of logically constrained term rewrite systems (LCTRSs, for short). 
However, the polynomial interpretation processor 
is not so effective against LCTRSs with bit-vector arithmetic (BV-LCTRSs, for short).
In this paper, 
we propose a novel DP processor for BV-LCTRSs to solve a singleton DP problem consisting of a dependency pair forming a self-loop.
The processor is based on an acyclic directed graph such that the nodes are bit-vectors and any dependency chain of the problem is projected to a path of the graph.
We show a sufficient condition for the existence of such an acyclic graph, and simplify it for a specific case.
\end{abstract}

\section{Introduction}
\label{sec:intro}
\emph{Logically constrained term rewrite systems} (LCTRSs, for short)~\cite{KN13frocos} 
are expected to be useful computational models for verifying not only functional but also imperative programs~\cite{FKN17tocl}.
Especially, \emph{LCTRSs with bit-vector arithmetic} (BV-LCTRSs, for short) are useful for programs written in C or other languages with \emph{fixed-width integers} such as \texttt{int} of C because primitive data types, structures, and unions are represented by bit-vectors in a natural and precise manner~\cite{KNS19ss}.
In proving validity of an equation w.r.t.\ a given rewrite system by means of \emph{rewriting induction}~\cite{Red90,FKN17tocl}, we need to \emph{frequently} try to prove termination 
of rewrite systems obtained by adding rewriting rules for induction hypotheses into the given system.
Therefore, for verification tools based on rewriting induction, the performance of proving termination of rewriting systems has a great influence on the proof power and execution time.

The \emph{dependency pair framework} (DP framework, for short)~\cite{GTS04} equipped with \emph{DP processors} which decompose DP problems is a well investigated technique for proving termination of rewrite systems, and has been extended to many kinds of constrained rewrite systems including LCTRSs \cite{FK08,FGPSF09,Kop13termination,SNSU18eptcs}.
Some fundamental DP processors are applicable to almost all kinds of rewrite systems without any change.
For example, the \emph{dependency graph processor} 
based on SCC decomposition
is applicable to LCTRSs.
On the other hand, the \emph{polynomial interpretation processor}, one of the most powerful processors in proving termination of LCTRSs with integer arithmetic, is applicable to a DP problem of BV-LCTRSs but not so effective against it:
It is ineffective if it contains a dependency pair with a usable rule for an operator of BVs such as addition, which may cause overflow and/or underflow.
To enhance the power of proving termination of BV-LCTRSs, we need to develop DP processors specific to BV-LCTRSs.

In this paper, we propose a novel DP processor, called a \emph{singleton self-loop removal processor} (SSR processor, for short), aiming at developing a method to prove termination of BV-LCTRSs.
Here, a dependency pair is said to \emph{form a self-loop} if it forms a dependency chain of length two or more, and a DP problem is called a \emph{singleton self-loop} problem if it is a singleton set, the pair in which forms a self-loop.
The processor takes a singleton self-loop DP problem as an input and is based on an acyclic directed graph such that the nodes are bit-vectors and any dependency chain of the problem is projected to a path of the graph.
We show a sufficient condition for the existence of such an acyclic graph, and simplify it for a specific case.
Note that the processor returns the empty set---the solved DP problem---if the sufficient condition is satisfied by a given singleton self-loop DP problem.

In the rest of the paper, familiarity with basic notions and notations on term rewriting~\cite{BN98,Ohl02} is assumed.
We follow the definition of LCTRSs in~\cite{KN13frocos,FKN17tocl}.
For brevity, we use the 4-bits bit-vectors for type {\ttfamily int}. 
We denote the set of bit-vectors of length $n$ by $\BV_n$.
To distinguish bit-vectors from decimal numbers, we follow the SMT-LIB %
notation for bit-vectors:
A bit-vector $c\in\BV_n$, which is written as a binary numeral in \textsf{sans-serif} font, is denoted by $\bv{\mathit{c}}$.
We often use regular expressions for binary numerals, e.g., $\symb{0}^3$ stands for $\symb{0}\symb{0}\symb{0}$.


\section{From C Programs to BV-LCTRSs}

A set $\cS$ of sorts for bit-vectors includes sort $bv_n$ for the $n$-bits bit-vectors ($n\geq 1$): $\cS\supseteq{\{bool\}\cup\{bv_n\mid n\geq1\}}$.
The set $\Valbv$ of values is $\{\symb{true},\symb{false}: bool\}\cup\bigcup_{n\geq1}\{b: bv_n \mid b\in\BV_n\}$.
The set $\Sigmabvtheory$ of theory symbols for bit-vectors is an extension of the \emph{core theory} $\Sigmacoretheory$~\cite{FKN17tocl} for logical connectives ($\lor,\land,\lnot$):
$\Sigmatheory = \Sigmacoretheory\cup\Val\cup
\{ {+_{bv_n}}: bv_n \times bv_n \Rightarrow bv_n, ~ {=_{bv_n}},{<_{bv_n,S}},{<_{bv_n,U}},{\geq_{bv_n,S}},{\geq_{bv_n,U}}:bv_n \times bv_n \Rightarrow bool, ~\ldots \mid n \geq 1\}$.
We drop the subscript $bv_n$ from $+_{bv_n}$, $<_{bv_n,S}$, and so on 
if it is clear from the context.
The interpretation of theory symbols for bit-vectors follow the usual semantics of bit-vector arithmetic~\cite{KS16}.
%

\begin{example}
\label{ex:bv-lctrs}
    The C program in Listing~\ref{list:cnt} is transformed into the following BV-LCTRS~\cite{KNS19ss}:
    \[
      \cRcnt\!=\!
      \left\{
      \begin{array}{@{}r@{\,}c@{\,}lc@{}}
        \symb{cnt}(x) & \to & \symb{u}_1(x,\bv{0000},\bv{0000}) & \\
        \symb{u}_1(x,i,z) & \to & \symb{u}_1(x,i\,{+}\,\bv{0001},z\,{+}\,\bv{0001}) & [i\,{<_S}\,x] \\
        \symb{u}_1(x,i,z) & \to & z & [i\,{\geq_S}\,x]
      \end{array}	
      \right\}
    \]
    where $\symb{cnt}:bv_{4}\Rightarrow bv_{4}$ and $\symb{u}_1:bv_{4} \times bv_{4} \times bv_{4} \Rightarrow bv_{4}$.
    Note that the above LCTRS is a simplified one by means of \emph{chaining}~(cf.~\cite[Section~7]{FK09}).
    Note also that \emph{calculation rules}~\cite{FKN17tocl} such as $x + y \to z ~ [ z = x + y ]$ are implicitly included in $\cRcnt$.
    For example, we have that 
    $
            \symb{cnt}(\bv{0010}) \to_{\cRcnt} \symb{u}_1(\bv{0010},\bv{0000},\bv{0000}) 
            \to_{\cRcnt} \symb{u}_1(\bv{0010},\bv{0000}+\bv{0001},\bv{0000}+\bv{0001}) 
            \to_{\cRcnt} \symb{u}_1(\bv{0010},\bv{0001},\bv{0000}+\bv{0001})
            \to_{\cRcnt} \cdots 
            \to_{\cRcnt} \bv{0010}
    $.
\end{example}

\begin{lstlisting}[caption={A C program defining a function to count $x$ times}, label=list:cnt, float=t, abovecaptionskip=\medskipamount]
int cnt(int x){
    int z=0;
    for(int i=0; i<x; i++) z++;
    return z;
}
\end{lstlisting}

\section{The DP Framework for LCTRSs}

The DP framework~\cite{GTS04} for TRSs has been extended for LCTRSs~\cite{Kop13termination}.

Let $\cR$ be an LCTRS.
The marked symbol of a defined symbol $f:\iota_1\times\cdots\times\iota_n\Rightarrow\iota\in\cD_\cR$ is denoted by $f^\#$ and the set of marked symbols for $\cD_\cR$ is denoted by $\cD^\#_\cR$.
We introduce a fresh basic sort $dpsort$, and $f^\#$ has sort $\iota_1\times\cdots\times\iota_n\Rightarrow dpsort$.
If $t=f(t_1,\ldots,t_n)$ with $f\in\cD_\cR$, then $f^\#(t_1,\ldots,t_n)$ is denoted by $t^\#$.
For each rule $\ell\to r ~ [\phi]\in\cR$, a constrained rewrite rule $\ell^\#\to t^\#\>[\phi]$ is called a \emph{dependency pair} (DP, for short) of $\cR$ if $t$ is a subterm of $r$ and $root(t)\in\cD_\cR$.
The set of DPs of $\cR$ is denoted by $\DP(\cR)$.
In the following, we use $\cP$ as a set of DPs of $\cR$, i.e., $\cP\subseteq \DP(\cR)$.
A sequence $\rho_1,\rho_2,\ldots$ of DPs in $\cP$ is called a \emph{dependency chain} of $\cP$ ($\cP$-chain, for short) 
if there are substitutions $\gamma_1,\gamma_2,\ldots$ such that for each $i > 0$, $\gamma_i$ \emph{respects} $\rho_i=(s_i^\#\to t_i^\#\>[\phi_i])$---$\Ran(\gamma_i|_{\Var(\phi_i)\cup(\Var(t_i)\setminus\Var(s_i))})\subseteq \Val$ and $[\![ \phi_i\gamma_i]\!]=\top$---and $t_i^\#\gamma_i\to_\cR^*s_{i+1}^\#\gamma_{i+1}$.  
%
A \emph{DP problem}  $(\cP,\cR)$, abbreviated to $\cP$, is called \emph{chain-free} if there is no infinite $\cP$-chain.
\begin{theorem}[\cite{Kop13termination}]
\label{termination}
An LCTRS $\cR$ is terminating iff the DP problem $\DP(\cR)$ is chain-free.
\end{theorem}

A DP processor $\Proc$ is a function that maps a DP problem to a finite set of DP problems: $\Proc(\cP) \subseteq 2^\cP$. 
We say that $\Proc$ is \emph{sound} if for any DP problem $\cP$, $\cP$ is chain-free, whenever all DP problems in $\Proc(\cP)$ are chain-free.
%
If the initial problem $\DP(\cR)$ is decomposed into the solved DP problem $\emptyset$ by applying sound DP processors, then the framework succeeds in proving termination of $\cR$.

A \emph{dependency graph} (DG, for short) of $\cP$ is a directed graph $\mathcal{G}=(\cP,\cE)$, denoted by $\DG(\cP)$, such that $\cE=\{(\rho_1,\rho_2) \mid \rho_1,\rho_2\in\cP, ~ \mbox{the seqeunce $\rho_1,\rho_2$ is a $\cP$-chain}\}$.
Moreover, a directed graph $\mathcal{G'}=(\cP,\cE')$ with $\cE'\supseteq \cE$ is called a \emph{DG approximation} of $\cP$.
A computation of DG approximations can be seen in~\cite{Kop13termination}.

\begin{theorem}[cf.~\cite{Kop13termination}]
The \emph{dependency graph processor} $\ProcSCC$ such that 
$\ProcSCC(\cP)=\{\cP' \mid$ $\cP'$ are the nodes of an SCC in a DG approximation of $\cP$
$\}$
is a sound 
DP processor.
\end{theorem}

\begin{example}
\label{ex:DP}
    Consider $\cRcnt$ in Example~\ref{ex:bv-lctrs} again.
    The following pairs are the DPs of $\cRcnt$:
    \[
      \DP(\cRcnt)=
      \left\{
      \begin{array}{@{\,}c@{~~}r@{\>}c@{\>}l@{\>}c@{\,}}
        (1) & \symb{cnt}^\#(x) & \to & \symb{u}^\#_1(x,\bv{0000},\bv{0000}) & \\
        (2) & \symb{u}^\#_1(x,i,z) & \to & \symb{u}^\#_1(x,i\,{+}\,\bv{0001},z\,{+}\,\bv{0001}) & [i\,{<_S}\,x]
      \end{array}	
      \right\}
    \]
    Since $\DG(\DP(\cRcnt))=(\DP(\cRcnt),\{ ((1),(2)), ((2),(2))\})$,
    we have that $\ProcSCC(\DP(\cRcnt)) = \{\>\{\,(2)\,\}\>\}$.
    In the following, we denote $\{\,(2)\,\}$ by $\cP_1$:
    \[
      \cP_1 =
      \{~
        (2) ~ \symb{u}^\#_1(x,i,z) \to \symb{u}^\#_1(x,i\,{+}\,\bv{0001},z\,{+}\,\bv{0001}) ~ [i\,{<_S}\,x]
      ~\}
    \]
\end{example}

\section{DP Processor for Singleton Self-Loop Removal}

The polynomial interpretation (PI, for short) processor over the integers (cf.~\cite[Theorem~10]{Kop13termination}) is one of the most powerful DP processors in proving termination of LCTRSs with integer arithmetic.
As indicated in Section~\ref{sec:intro}, however, the PI processor is ineffective against DP problems of BV-LCTRSs in the case where a usable calculation rule may cause overflow and/or underflow, because such a rule (e.g., $x + y \to z ~ [\, z = x + y\,]$) cannot be ordered by any meaningful PI order. 
In this section, we propose a DP processor, called an \emph{singleton self-loop removal processor}, that solves singleton self-loop DP problems under a certain condition.

A \emph{singleton self-loop DP problem} in this paper is assumed to be a singleton set of the form 
$\{\>\symb{f}^\#(x_1,\ldots,x_n) \to \symb{f}^\#(t_1,\ldots,t_n) ~ [\phi]\>\}$
that forms a chain of length two or more.
For example, $\cP_1$ in Example~\ref{ex:DP} is a singleton self-loop DP problem.
In the rest of this section, we let $\cP = \{\>f^\#(x_1,\ldots,x_n) \to f^\#(t_1,\ldots,t_n) ~ [\phi]\>\}$ be a singleton self-loop DP problem, where $f$ has sort $\iota_1\times\cdots\times\iota_{i-1}\times bv_l\times\iota_{i+1}\times\cdots\times\iota_n \Rightarrow \iota$, $l$ is a natural number, $x_1,\ldots,x_n$ are pairwise distinct variables, and $\phi$ is satisfiable.
    
In rewriting a term by the DP~(2) of $\cP_1$, the first and third arguments of $\symb{u}^\#_1$ do not affect the constraint $i <_S x$, i.e., they preserve the evaluation of the constraint, while the second argument of $\symb{u}^\#_1$ does not preserve the value of $i <_S x$---for a substitution $\theta$ such that $(i <_S x)\theta$ holds, $(i <_S x)\{i\mapsto i+\symb{1}\}\theta$ ($=(i+\symb{1} <_S x)\theta$) may not hold.
We formulate this notion as follows:
A rewrite rule $f^\#(x_1,\ldots,x_n) \to f^\#(t_1,\ldots,t_n) ~[\phi]$ is said to \emph{preserve its constraint w.r.t.\ $\bar{I}$} ($\subseteq \{1,\ldots,n\}$) if all of the following hold:
\begin{bracketenumerate}
    \item 
    $\{x_i \mid 1 \leq i \leq n, ~ i \notin \bar{I} \} \subseteq \Var(\phi)$,
    \item 
    for each $j\in \{1,\ldots,n\}$, if $x_j\in\Var(\phi)$, then $t_j\in\cT(\Sigmabvtheory,\cV)$,
        and
   \item 
    $(\exists y_1,\ldots,y_m.\ \phi) \mathrel{\Leftrightarrow} (\exists y_1,\ldots,y_m.\ \phi)\theta$ is valid, where $\{y_1,\ldots,y_m\} = \Var(\phi) \setminus \{x_1,\ldots,x_n\}$ and $\theta=\{ x_j\mapsto t_j \mid 1 \leq j \leq n, ~ j \in \bar{I}  \}$.
\end{bracketenumerate}
For example, the DP (2) in $\cP_1$ preserves its constraint w.r.t.\ $\{1,3\}$.

To prove chain-freeness of $\cP_1$, we use the fact that, given a DP problem $\cP$, if there exists an acyclic directed graph such that the nodes are bit-vectors and any $\cP$-chain is projected to a path of the graph, then $\cP$ is solved, i.e., $\cP$ is chain-free.

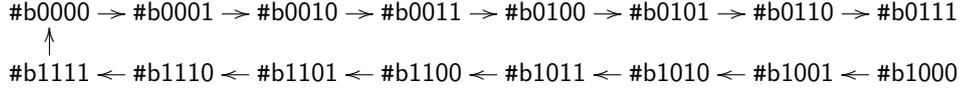
\begin{figure}[tb]
\vspace{-12pt}
    \[
        \xymatrix@=10pt{
            \bv{0000} \ar[r] & \bv{0001} \ar[r] & \bv{0010} \ar[r] & \bv{0011} \ar[r] & \bv{0100} \ar[r] & \bv{0101} \ar[r] & \bv{0110} \ar[r] & \bv{0111} \\
            \bv{1111} \ar[u] & \bv{1110} \ar[l] & \bv{1101} \ar[l] & \bv{1100} \ar[l] & \bv{1011} \ar[l] & \bv{1010} \ar[l] & \bv{1001} \ar[l] & \bv{1000} \ar[l]
        }
    \]
\vspace{-8pt}
    \caption{The acyclic graph obtained from $\symb{u}^\#_1(t_1,b,t_3)$ by projecting to the second argument.}
    \label{fig:graph}
\end{figure}

Let us consider the DP~(2) in $\cP_1$.
Let $\pi_1$ be a projection of terms rooted by $\symb{u}^\#_1$ to the second argument of the root symbol, 
i.e., $\pi_1(\symb{u}^\#_1(t_1,t_2,t_3))=t_2$.
Then, we construct a directed graph such that the nodes are the 4-bits bit-vectors and the edges illustrated in Figure~\ref{fig:graph} are obtained from $\cP_1$ by applying $\pi_1$ to ground instances of the DP~(2) in $\cP_1$.
Since the graph is acyclic and any $\cP_1$-chain is projected to a path of the graph, the graph ensures the non-existence of infinite $\cP_1$-chains.

In the following, we show a sufficient condition for the existence of such an acyclic graph.
\begin{restatable}{theorem}{maintheorem}
\label{teiri2}
    Let $i \in \{1,\ldots,n\}$.
    Suppose that 
    \begin{bracketenumerate}
    \setcounter{enumi}{3}
        \item 
$f^\#(x_1,\ldots,x_n) \to f^\#(t_1,\ldots,t_n) ~ [\phi]$ preserves its constraint w.r.t.\ $\{1,\ldots,i-1,i+1,\ldots,n\}$, 
        \item 
there exists $a \in \{0,\ldots,l\}$ such that 
    $t_i-x_i =_{bv_l} \bv{\mathit{c}\symb{1}0^{\mathit{a}}}$ is valid
    for some $c\in\{\symb{0},\symb{1}\}^{l-a-1}$,
            and
        \item 
there exists some terms $u,v:bv_l\in\cT(\Sigmabvtheory,\Var(\phi))$ such that 
    \begin{enumerate}
    \renewcommand{\labelenumi}{\alph{enumi}.}
        \item $\phi \mathrel{\Rightarrow} (u =_{bv_l} u\theta \land v =_{bv_l} v\theta
        \land v-u\geq_U \bv{0^{\mathit{l-a-\mathrm{1}}}10^{\mathit{a}}})$ is valid,
        where $\theta=\{ x_j\mapsto t_j \mid 1 \leq j \leq n\}$,
        and
        \item 
        $(\forall x_i.\ (\phi \mathrel{\Rightarrow} ((x_i<_Uu \lor v\leq_Ux_i) \land u<_Uv)))
        \lor
        (\forall x_i.\ (\phi \mathrel{\Rightarrow} (v\leq_Ux_i \land x_i <_Uu)))$
        is valid.
    \end{enumerate}
    \end{bracketenumerate}
    Then, $\cP$ is chain-free. 
\end{restatable}

The assumptions in Theorem~\ref{teiri2} mean the following, respectively:
\Enumi{(4)}
The evaluation of $\phi$ is only affected by the $i$-th argument of $f$ in applying the DP to terms;
\Enumi{(5)}
the $i$-th argument of $f$ plays a role of a \emph{loop variable}, and $t_i-x_i$ is a fixed amount ($\bv{\mathit{c}\symb{1}0^{\mathit{a}}}$) of the increment or decrement of the $i$-th argument at the application of the DP;
\Enumi{(6)}
the terms $u,v$ imply a fixed interval $[u,v)$ that is not affected by the application of the DP and has the length more than $\bv{0^{\mathit{l-a-\mathrm{1}}}10^{\mathit{a}}}$.
Note that if $v <_U u$ holds, then the interval is $\{ b \in \BV_l \mid b <_U v \lor u \leq_U b \}$.
In applying the DP to a term (i.e., $\phi$ is satisfied), the value of $x_i$ is out of the interval.
In other words, if the value of $x_i$ is in the interval, then $\phi$ is not satisfied and thus, the DP is not applicable.
By repeating the application of the DP, the value of $x_i$ always enter the interval.
This means that a $\cP$-chain can no longer be extended and thus, there is no infinite $\cP$-chain.
\begin{example}
Let us consider the DP (2) in $\cP_1$ again.
Regarding the second argument, the DP (2) satisfies both~\Enumi{(4)} and~\Enumi{(5)}:
$(i+\bv{0001}) - i = \bv{0001}$.
Let $u=x+\bv{1000}$ and $v=x+\bv{1001}$.
Then, $u,v$ satisfy both~\Enumi{(6)\,a} and~\Enumi{(6)\,b}:
$i <_S x \Rightarrow ({x+\bv{1000} =_{bv_4} x+\bv{1000}} \land {x+\bv{1001} =_{bv_4} x+\bv{1001}} \land {(x+\bv{1001})-x =_{bv_4} \bv{0001}}$ 
and
$(\forall i.\ (i <_S x \mathrel{\Rightarrow} ((i<_U x+\bv{1000} \lor x+\bv{1001}\leq_U i) \land x+\bv{1000} <_U x+\bv{1001})))
\lor
(\forall i.\ (i <_S x \mathrel{\Rightarrow} (x+\bv{1001}\leq_U i \land i <_U x+\bv{1000})))$
are valid.
Therefore, by Theorem~\ref{teiri2}, $\cP_1$ is chain-free, i.e., $\cRcnt$ is terminating.
\end{example}


In the assumption~\Enumi{(6)\,b} of Theorem~\ref{teiri2}, theory terms with sort $bv_l$ are interpreted as unsigned integers, but this does not mean that Theorem~\ref{teiri2} only works for BV-LCTRSs obtained from C programs where all \texttt{int} variables are unsigned.
For example, the variable \texttt{x}, \texttt{i}, and \texttt{z} in Listing~\ref{list:cnt} are signed ones.

To prove chain-freeness of $\cP$ by Theorem~\ref{teiri2}, we need to find terms $u,v$ that satisfy~\Enumi{(6)\,a} and~\Enumi{(6)\,b}.
It is not easy to find such terms $u,v$ mechanically, because each of them may be either a variable such as $x$ or a bit-vector expression consisting of variables and constants such as $x+\bv{0111}$.
To overcome this difficulty, we propose another implementable criterion that can be applied in a specific case.

Let us focus on the case where $a=0$ (i.e., increment or decrement of the $i$-th argument is an odd number) in Theorem~\ref{teiri2}.
Since
$\bv{0^{\mathit{l-a-\mathrm{1}}}10^\mathit{a}} = \bv{0^{\mathit{l-\mathrm{1}}}1}$,
the formula $v-u\geq_U 
\bv{0^{\mathit{l-a-\mathrm{1}}}10^\mathit{a}}
$
is equivalent to $u \ne_{bv_l} v$.
If~\Enumi{(6)\,b} holds, then $\phi \mathrel{\Rightarrow} (u \ne_{bv_l} v)$ is valid.
In addition, neither $\forall x_i.\ ((x_i<_Uu \lor v\leq_Ux_i) \land u<_Uv)$ nor $\forall x_i.\ (v\leq_Ux_i \land x_i<_U u)$ is satisfiable.
Thus,~\Enumi{(6)\,b} is equivalent to
unsatisfiability of $\forall x_i.\ \phi$ which does not contain either $u$ or $v$ and enables us to drop~\Enumi{(6)\,a}.
In summary, Theorem~\ref{teiri2} is simplified in the case where $a=0$.
\begin{theorem}
\label{teiri3}
    Let $i \in \{1,\ldots,n\}$.
    Suppose that 
    \begin{bracketenumerate}
    \setcounter{enumi}{3}
    \leftskip=1ex
        \item 
        $f^\#(x_1,\ldots,x_n) \to f^\#(t_1,\ldots,t_n) ~ [\phi]$ preserves its constraint w.r.t.\ $\{1,\ldots,i-1,i+1,\ldots,n\}$, 
        \item[\Enumi{(5')}]  
        $t_i - x_i =_{bl_l} \bv{\mathit{c}\symb{1}}$ is valid
        for some $c\in\{\symb{0},\symb{1}\}^{l-1}$,
            and
        \item[\Enumi{(6')}]
        $\forall x_i.\ \phi$ is unsatisfiable.
    \end{bracketenumerate}
    Then, $\cP$ is chain-free. 
\end{theorem}

Theorem~\ref{teiri3} can be applied only in the case when $a=0$ in Theorem~\ref{teiri2}, but in most practical programs, loop variables are incremented by one.
Thus, Theorem~\ref{teiri3} must be sufficient to prove termination of BV-LCTRSs obtained from such programs.
Besides, 
the criterion in Theorem~\ref{teiri3} is more implementable than that in Theorem~\ref{teiri2}.

\begin{example}
    Since $\forall i .\ i<_Sx$ is unsatisfiable, 
    by Theorem~\ref{teiri3}, $\cP_1$ is chain-free.
\end{example}

Finally, we propose a DP processor based on Theorems~\ref{teiri2} and~\ref{teiri3}.
\begin{definition}
    Suppose that $\cP$ satisfies the assumptions in Theorems~\ref{teiri2} or~\ref{teiri3}.
    Then, given $\cP$, the \emph{singleton self-looping removal processor} $\ProcLoop$ returns $\{\emptyset\}$:
    $\ProcLoop(\cP)=\{\emptyset\}$.
\end{definition}
By Theorems~\ref{teiri2} and~\ref{teiri3}, it is clear that $\ProcLoop$ is a sound DP processor.


\section{Future Work}
\label{sec:future-work}


The applicability of the SSR processor $\ProcLoop$ is very limited because it works for singleton (self-loop) problems only.
We may make a single loop formed by two or more DPs a self-loop formed by a DP by means of chaining.
On the other hand, we need a device for multiple loops.
An idea for such loops is to extract an innermost loop, decomposing a multiple loop into an innermost one and the others;
for the latter, we overapproximate the innermost loop by replacing the innermost loop-variable and accumulators by fresh variables.
Our future work is to formulate and implement this idea.

In~\cite{FKS12,HGFS18}, bit-vectors and their operators are represented over integer arithmetic, e.g., by case analysis for the finite interval of integers ($[-2^{31},2^{31}-1]$) or \emph{modulo} relations.
In~\cite{CDKSW18}, bit-precise termination is synthesised over lexicographic linear ranking function templates.
We have to compare our method with such approaches from the theoretical and empirical points of view.




\bibliography{biblio}

\end{document}

%% file: mymacro.tex
\usepackage[all]{xy}

\def\cD{\mathcal{D}}
\def\cE{\mathcal{E}}
\def\cR{\mathcal{R}}
\def\cP{\mathcal{P}}
\def\cV{\mathcal{V}}
\def\cS{\mathcal{S}}
\def\cT{\mathcal{T}}

\def\symb#1{\mathsf{#1}}
\newcommand{\bv}[1]{{\ttsharp\symb{b}\symb{#1}}}

\def\BV{\mathbb{BV}}

\newcommand{\Sigmatheory}[1][]{\Sigma^{\mathit{#1}}_{\mathit{theory}}}
\def\Sigmacoretheory{\Sigmatheory[core]}

\def\Sigmabvtheory{\Sigmatheory}

\def\Var{{\mathcal{V}\mathit{ar}}}

\def\Dom{{\mathcal{D}\mathit{om}}}
\def\Ran{{\mathcal{R}\mathit{an}}}
\def\VRan{{\mathcal{VR}\mathit{an}}}

\newcommand{\Val}[1][]{{\mathcal{V}\!\mathit{al}}^{\mathit{#1}}}
\def\Valbv{\Val}

\newcommand{\ttsym}[1]{\mbox{\tt #1}}

\def\ttsharp{\ttsym{\symbol{35}}} 

\def\DP{\mathit{DP}}
\def\DG{\mathit{DG}}

\newcommand{\Proc}[1][]{\mathit{Proc}_{\mathit{#1}}}
\def\ProcSCC{\Proc[DG]}

\def\ProcLoop{\Proc[SSR]}

\newcommand{\Enumi}[1]{\textcolor{lipicsGray}{\sffamily\bfseries\upshape\mathversion{bold}#1}}

\def\cRcnt{\cR_1}